\documentclass{article}
\usepackage[cp1251]{inputenc}
\usepackage[english]{babel}
\usepackage{graphics} 
\usepackage{epsfig}
\usepackage{amsmath, amssymb}
\input{trasp_my.sty}
\usepackage{axodraw}
\def\footauthor{T. Shishkina, I. Sotsky}
\def\foottitle{Production of four leptons in $\gamma\gamma$-- interactions}

\begin{document}

\vspace{-6.5cm}

\title{\bf Helicity amplitude method for two electron positron pairs photoproduction.}
\vspace{0.3cm}
\author{T.Shishkina\\ shishkina@hep.by \\ \it  BSU, Minsk, Belarus \\  I. Sotsky \\ (sotsky@hep.by) \\ \it  BSU, Minsk, Belarus}
\date{}
\maketitle \thispagestyle{empty} \vspace{-0.8cm}
\begin{abstract}
\begin{center}
\begin{flushleft}
We present the matrix element  of $\gamma\gamma\rightarrow 4l$
process with four charged particles in final state. The
constructions are performed in frame of Standard Model   using the
helicity amplitude method.  Every \par \noindent possible
polarization state
  of initial and final   particles are considered.
\end{flushleft}
\end{center}
\end{abstract}

\vspace{-0.5cm}
 \section { Introduction} \vspace{0.3cm}
\large

$\;\;\;$In designed and constructed future linear colliders
besides of  $e^-e^-$ and $e^+e^-$ interactions the realization
$\gamma\gamma$ and $\gamma\,e$ modes are planned. In the  last
case high energy photon beams are created by Compton
backscattering of initial photons on high energy electrons. This
possibility will allow a detailed study of non-abelian nature of
electroweak interaction,  gauge boson coupling as well as the
couplings of gauge bosons with Higgs particles if it is light
enough to be produced. Since $W^{\pm}$ and Higgs bosons decay
within detector they can be obtained via their decay products, for
instance four leptons in final state.

 For successful  realization of such kind of experiments  an exact
calculation  of all backgrounds  and luminosity  of initial beams
value is required. These  data can be obtained using consideration
of $\gamma\gamma\;\rightarrow 4l$, $\gamma\gamma\;\rightarrow 2l$
and $\gamma\gamma\;\rightarrow 2l\; + \;photons\;\;$ processes.

Total cross sections of such  interactions  have been already
calculated and analyzed in refs. \cite{c1}-\cite{c3} about 30
years ago  and were found to be large enough:

\large
 $$\sigma = 6500nb\;\;\;\;(\gamma\gamma
\rightarrow 2e^-2e^+),$$ $$\sigma = 5.7nb\;\;\;\;(\gamma\gamma
\rightarrow e^+e^-\mu^+\mu^-),$$ $$\sigma =
0.16nb\;\;\;\;(\gamma\gamma \rightarrow 2\mu^-2\mu^+).$$

\noindent
 However these calculations have  used the low energy approximation, and
 obtained results are not applicable to analyze the result of   high energy experiments.

The matrix element of $\gamma\gamma\;\rightarrow 4l$ process has
been constructed also in ref. \cite{c4}. It was done in implicit
form at limit of small polar angles of pair productions. However
at that paper neither calculation of cross section no numerical
analyze are present. So one could not perform any numerical
congruence.

Besides that, processes of $\gamma\gamma$ scattering have
considered in ref. \cite{c5}. There was applied the algorithm
ALPHA for automatic computations of scattering amplitude. However
modern high energy experiments require calculation of the cross
section at definite polarization states of initial and final
particles that ALPHA method couldn't provide.

This process was also analysed in ref. \cite{c6}, where  several
numerical calculations were performed, and the dependence of total
cross section from  the energy of initial beam was investigated.

 Present paper is devoted to  construction of the matrix element
of $\gamma\gamma$ electroweak interaction with production of  four
leptons  using helicity amplitude method \cite{c7} -\cite{c10}.
Numerical integration of obtained amplitude under
 kinematics of future projects will be described in the next
paper.

\npage \large

 \vspace{1cm}
 \section{Construction and calculations} \vspace{0.5cm} \large

$\;\;\;$There are six topographically different Feynman diagrams
of electroweak interaction describing process
$\gamma\gamma\;\rightarrow 4l$ (see fig.1). Using C-, P- and
crossing symmetries one can build 40 different diagrams.
\vspace{3.4cm}

\begin{center}
\begin{picture}(400,100)
\Photon(10,180)(35,167){1}{2} \Photon(10,100)(35,114){1}{2}
\ArrowLine(35,167)(85,175) \ArrowLine(85,106)(35,114)
\ArrowLine(45,150)(35,167) \ArrowLine(35,114)(45,130)
\Photon(45,150)(45,130){1}{2} \ArrowLine(85,150)(45,150)
\ArrowLine(45,130)(85,130) \Text(15,185)[l]{\small $\gamma(k_1)$}
\Text(2,115)[l]{\small $\gamma(k_2)$}\Text(90,177)[l]{\small
$p_1$}\Text(90,150)[l]{\small $p_2$} \Text(90,130)[l]{\small
$p_3$}\Text(90,104)[l]{\small $p_4$} \Text(35,95)[l]{\small
$(1)$}\Text(-3,141)[l]{\small
$W,Z,\gamma$}\Vertex(35,167){2}\Vertex(35,114){2}\Vertex(45,150){2}
\Vertex(45,130){2}

\Photon(150,180)(175,167){1}{2} \Photon(150,100)(175,114){1}{2}
\ArrowLine(175,167)(225,175) \ArrowLine(225,106)(175,114)
\ArrowLine(185,140)(175,167) \ArrowLine(175,114)(185,140)
\Photon(185,140)(205,140){1}{2} \ArrowLine(225,130)(205,140)
\ArrowLine(205,140)(225,150) \Text(155,185)[l]{\small
$\gamma(k_1)$} \Text(150,117)[l]{\small
$\gamma(k_2)$}\Text(230,177)[l]{\small
$p_1$}\Text(230,150)[l]{\small $p_3$} \Text(230,130)[l]{\small
$p_4$}\Text(230,104)[l]{\small $p_2$} \Text(175,95)[l]{\small
$(2)$}\Text(183,157)[l]{\small
$W,Z,\gamma$}\Vertex(175,167){2}\Vertex(175,114){2}\Vertex(205,140){2}
\Vertex(185,140){2}

\Photon(290,180)(315,167){1}{2} \Photon(290,100)(315,114){1}{2}
\ArrowLine(315,167)(345,159) \ArrowLine(365,106)(315,114)
\ArrowLine(315,114)(315,167) \ArrowLine(345,159)(365,143)
\Photon(345,159)(355,169){1}{2} \ArrowLine(365,158)(355,169)
\ArrowLine(355,169)(365,180) \Text(290,185)[l]{\small
$\gamma(k_1)$} \Text(290,117)[l]{\small
$\gamma(k_2)$}\Text(370,185)[l]{\small
$p_1$}\Text(370,160)[l]{\small $p_2$} \Text(370,145)[l]{\small
$p_3$}\Text(370,104)[l]{\small $p_4$} \Text(315,95)[l]{\small
$(3)$}\Text(315,180)[l]{\small
$W,Z,\gamma$}\Vertex(315,167){2}\Vertex(315,114){2}\Vertex(345,159){2}
\Vertex(355,169){2}

\end{picture}
\end{center}

\vspace{0cm}

\begin{center}
\begin{picture}(400,100)
\Photon(10,180)(50,140){1}{4} \Photon(10,100)(50,140){1}{4}
\ArrowLine(65,155)(85,175) \ArrowLine(85,106)(65,125)
\Photon(50,140)(65,155){1}{2}\Photon(50,140)(65,125){1}{2}
\ArrowLine(85,150)(65,155) \ArrowLine(65,125)(85,130)
\Text(15,185)[l]{\small $\gamma(k_1)$} \Text(1,117)[l]{\small
$\gamma(k_2)$}\Text(90,177)[l]{\small
$p_1$}\Text(90,150)[l]{\small $p_2$} \Text(90,130)[l]{\small
$p_3$}\Text(90,104)[l]{\small $p_4$} \Text(35,95)[l]{\small
$(4)$}\Text(48,155)[l]{\small $W$}\Text(48,125)[l]{\small
$W$}\Vertex(50,140){2}\Vertex(65,155){2}\Vertex(65,125){2}

\Photon(150,180)(175,160){1}{3} \Photon(150,100)(175,120){1}{3}
\ArrowLine(200,160)(225,175) \ArrowLine(225,106)(200,120)
\Photon(175,120)(175,160){1}{3}
\Photon(175,120)(200,120){1}{2}\Photon(175,160)(200,160){1}{2}
\ArrowLine(200,120)(225,130) \ArrowLine(225,150)(200,160)
\Text(155,185)[l]{\small $\gamma(k_1)$} \Text(140,110)[l]{\small
$\gamma(k_2)$}\Text(230,177)[l]{\small
$p_1$}\Text(230,150)[l]{\small $p_2$} \Text(230,130)[l]{\small
$p_3$}\Text(230,104)[l]{\small $p_4$} \Text(175,95)[l]{\small
$(5)$}\Text(183,168)[l]{\small $W$}\Text(183,130)[l]{\small
$W$}\Text(183,168)[l]{\small $W$} \Text(160,142)[l]{\small $W$}
\Vertex(175,160){2}\Vertex(175,120){2}\Vertex(200,120){2}
\Vertex(200,160){2}

\Photon(290,180)(315,167){1}{2} \Photon(290,100)(315,114){1}{2}
\Photon(315,167)(350,167){1}{3} \ArrowLine(365,106)(315,114)
\ArrowLine(315,114)(335,134) \ArrowLine(335,134)(365,134)
\Photon(315,167)(335,134){1}{3} \ArrowLine(365,158)(350,167)
\ArrowLine(350,167)(365,180) \Text(295,185)[l]{\small
$\gamma(k_1)$} \Text(279,110)[l]{\small
$\gamma(k_2)$}\Text(370,185)[l]{\small
$p_1$}\Text(370,160)[l]{\small $p_2$} \Text(370,135)[l]{\small
$p_3$}\Text(370,104)[l]{\small $p_4$} \Text(315,95)[l]{\small
$(6)$}\Text(332,175)[l]{\small $W$}\Text(310,145)[l]{\small
$W$}\Vertex(315,167){2}\Vertex(315,114){2}\Vertex(350,167){2}
\Vertex(335,134){2} \Text(170,75)[l]{ \Large$Fig.1.$}
\end{picture}
\end{center}
\vspace{-2cm} \large

The diagrams  containing charged current exchange are excluded
because only process with four charged leptons in final state is
considered. Matrix elements for  remaining diagrams (1)-(3) have
the following form:  \vspace{0.5cm}

 \large

\begin{eqnarray}\label{c1}
\begin{array}{c}
 M_1 =
\frac{{\displaystyle-ie^4}}{\displaystyle(k_1-p_1-p_2)^2}\overline{u}(p_1)\widehat{\varepsilon}(k_1)\frac{\displaystyle\widehat{p}_1-\widehat{k}_1+m}{\displaystyle(p_1-k_1)^2-m^2}\gamma^{\mu}v(p_2)\overline{u}(p_3)\gamma_{\mu}\frac{\displaystyle\widehat{k}_2-\widehat{p}_4+m}{\displaystyle(k_2-p_4)^2-m^2}\widehat{\varepsilon}(k_2)v(p_4)-
\\
{ie^2\left(\frac{\displaystyle g}{\displaystyle
2cos(\theta_W)}\right)^2}{D_{\mu\nu}(k_1-p_1-p_2)}\overline{u}(p_1)\widehat{\varepsilon}(k_1)\frac{\displaystyle\widehat{p}_1-\widehat{k}_1+m}{\displaystyle(p_1-k_1)^2-m^2}\gamma^{\mu}(g_V+g_A\gamma_5)v(p_2)\times
\\
\overline{u}(p_3)\gamma^{\nu}(g_V+g_A\gamma_5)\frac{\displaystyle\widehat{k}_2-\widehat{p}_4+m}{\displaystyle(k_2-p_4)^2-m^2}\widehat{\varepsilon}(k_2)v(p_4),

\\ \\

M_2 =
\frac{{\displaystyle-ie^4}}{\displaystyle(p_3+p_4)^2}\overline{u}(p_1)\widehat{\varepsilon}(k_1)\frac{\displaystyle\widehat{p}_1-\widehat{k}_1+m}{\displaystyle(p_1-k_1)^2-m^2}\gamma^{\mu}\frac{\displaystyle\widehat{k}_2-\widehat{p}_2+m}{\displaystyle(k_2-p_2)^2-m^2}\widehat{\varepsilon}(k_2)v(p_2)\overline{u}(p_3)\gamma_{\mu}v(p_4)-
\\
{ie^2\left(\frac{\displaystyle g}{\displaystyle
2cos(\theta_W)}\right)^2}{D_{\mu\nu}(p_3+p_4)}\overline{u}(p_1)\widehat{\varepsilon}(k_1)\frac{\displaystyle\widehat{p}_1-\widehat{k}_1+m}
{\displaystyle(p_1-k_1)^2-m^2}\gamma^{\mu}(g_V+g_A\gamma_5)\times
\\
\frac{\displaystyle\widehat{k}_2-\widehat{p}_2+m}{\displaystyle(k_2-p_2)^2-m^2}\widehat{\varepsilon}(k_2)v(p_2)\overline{u}(p_3)\gamma^{\nu}(g_V+g_A\gamma_5)v(p_4),

\\ \\
 M_3
=\frac{{\displaystyle-ie^4}}{\displaystyle(p_1+p_2)^2}\overline{u}(p_3)\gamma^{\mu}\frac{\displaystyle\widehat{p}_1+\widehat{p}_2+\widehat{p}_3+m}{\displaystyle(p_1+p_2+p_3)^2-m^2}\widehat{\varepsilon}(k_1)\frac{\displaystyle\widehat{k}_2-\widehat{p}_4+m}{\displaystyle(k_2-p_4)^2-m^2}\widehat{\varepsilon}(k_2)v(p_4)\overline{u}(p_1)\gamma_{\mu}v(p_2)-
\\ {ie^2\left(\frac{\displaystyle g}{\displaystyle
2cos(\theta_W)}\right)^2}{D_{\mu\nu}(p_1+p_2)}\overline{u}(p_3)\gamma^{\mu}(g_V+g_A\gamma_5)\frac{\displaystyle\widehat{p}_1+\widehat{p}_2+\widehat{p}_3+m}{\displaystyle(p_1+p_2+p_3)^2-m^2}\widehat{\varepsilon}(k_1)\times
\\
\frac{\displaystyle\widehat{k}_2-\widehat{p}_4+m}{\displaystyle(k_2-p_4)^2-m^2}\widehat{\varepsilon}(k_2)v(p_4)\overline{u}(p_1)\gamma^{\nu}(g_V+g_A\gamma_5)v(p_2).
\end{array}
\end{eqnarray}

\npage \large

\vspace{0.5cm} \noindent Here $\widehat{p}_1 =
p_1^{\mu}\gamma_{\mu}$, where $p_1^{\mu}$ is $\mu$ - component of
four momentum $p_1$;
$\widehat{\varepsilon}(k_1)=\varepsilon^{\mu}(k_1)\gamma_{\mu}$,
where $\varepsilon^{\mu}(k_1)$ --  $\mu$ - component of
polarization vector of photon with four momentum $k_1$,
$D_{\mu\nu}(q)$ -- propagator of $Z^0$-- boson with momentum $q$.

 \large  Corresponding cross section has the
form:
\begin{eqnarray}\label{c2}
\begin{array}{c}

\sigma = \frac{\displaystyle 1}{\displaystyle
4(k_1k_2)}\int|M|^2d\Gamma, \large
\end{array}
\end{eqnarray}
\noindent where $$d\Gamma
=
\frac{d^3p_1}{(2\pi)^32p_1^0}\frac{d^3p_2}{(2\pi)^32p_2^0}\frac{d^3p_3}{(2\pi)^32p_3^0}\frac{d^3p_4}{(2\pi)^32p_4^0}(2\pi)^4\delta(k_1+k_2-p_1-p_2-p_3-p_4)$$

\noindent is phase space element.
\par In this paper squared  matrix elements  are
constructed  using helicity amplitude  method (see, for example,
refs. \cite{c7}-\cite{c10}). It allows to calculate  cross section
directly for each definite polarization state of initial and final
particles. Matrix element constructed  by this method consists of
 invariants  without any bispinor, so  many difficulties are
excluded in squaring and numerical integrating.

We present here the final form of matrix elements of the diagrams
(1) - (3) (see fig.1.)  in case of electromagnetic $\gamma\gamma-$
interaction only.

Matrix elements (\ref{c3}) - (\ref{c6}) correspond to  the first
diagram and the main polarization states. Whole set of
rearrangements of initial and final particles are considered in
every equation (\ref{c3}) - (\ref{c12}).

 {\large

\begin{eqnarray}\label{c3}
\begin{array}{c}

M_1(+,-,+,-,+,-)=\,\left\{-4{(2e)}^4\,D(k_3^2)\,N_1(p_1,p_2)N_2(p_3,p_4)\Bigr[2\sqrt{(k_1p_3)(k_1p_2)}\times\right.
\\
\left.
\sqrt{(p_1p_2)(p_1p_3)}e^{i(\Delta\varphi_1)}-(p_1p_2)(p_1p_3)-(k_1p_3)(k_1p_2)e^{i(\Delta\varphi_2)}\Bigl]\sqrt{(p_1p_2)(p_3p_4)}\right\}-
\\ \Bigl\{{(p_1\leftrightarrow p_3)}\Bigr\}
e^{i(\Delta\varphi_3)}-\Bigl\{{(p_2\leftrightarrow p_4)}\Bigr\}
e^{i(\Delta\varphi_4)}+\Bigl\{{(p_1\leftrightarrow
p_3,p_2\leftrightarrow p_4)}\Bigr\} e^{i(\Delta\varphi_{1+-})},

\end{array}
\end{eqnarray}
\vspace{0.2cm}

\begin{eqnarray}\label{c6}
\begin{array}{c}

M_1(+,-,-,+,+,-)=\,\left\{4{(2e)}^4\,D(k_3^2)\,N_1(p_1,p_2)N_2(p_3,p_4)\Bigr[2\sqrt{(k_1p_3)(k_1p_1)}\times\right.
\\
\left.
\sqrt{(p_1p_2)(p_3p_2)}e^{i(\Delta\varphi_5)}-(p_1p_2)(p_3p_2)-(k_1p_3)(k_1p_1)e^{i(\Delta\varphi_6)}\Bigl]\sqrt{(p_1p_2)(p_3p_4)}\right\}+
\\ \Bigl\{{(k_1\leftrightarrow
k_2,p_1\leftrightarrow p_2,p_3\leftrightarrow p_4)}\Bigr\}^{*}
e^{i(\Delta\varphi_{7})},

\end{array}
\end{eqnarray}

\begin{eqnarray}\label{c5}
\begin{array}{c}

M_1(+,+,+,-,+,-)=\,{\Bigl\{4{(2e)}^4\,D(k_3^2)\,N_1(p_1,p_2)N_2(p_3,p_4)}\Bigl[(p_1k_1)+(p_2k_1)-(p_1p_2)\Bigr]\times\\
(p_2p_4)\sqrt{(p_1p_2)(p_3p_4)}\Bigr\}+\Bigl\{{(k_1\leftrightarrow
k_2)}\Bigr\}
e^{i(\Delta\varphi_{1+,+})}-\Bigl\{{(p_1\leftrightarrow
p_3)}\Bigr\} e^{i(\Delta\varphi_8)}- \\
\Bigl\{{(p_1\leftrightarrow p_3,k_1\leftrightarrow k_2)}\Bigr\}
e^{i(\Delta\varphi_{8})}e^{i(\Delta\varphi_{1++})},

\end{array}
\end{eqnarray}

\begin{eqnarray}\label{c4}
\begin{array}{c}

M_1(+,+,-,+,+,-)=\,{\Bigl\{4{(2e)}^4\,D(k_3^2)\,N_1(p_1,p_2)N_2(p_3,p_4)}\Bigl[(k_1p_1)+(k_1p_2)-(p_1p_2)\Bigr]\times\\
(p_1p_4)\sqrt{(p_1p_2)(p_3p_4)}\Bigr\}+\Bigl\{{(k_1\leftrightarrow
k_2)}\Bigr\} e^{i(\Delta\varphi_{1++})}.
\end{array}
\end{eqnarray}
\vspace{0.2cm} \vspace{0.2cm}

\vspace{1cm}

 \large Equations (\ref{c7}) and (\ref{c8}) describe the second diagram:

\npage
 {\large
\begin{eqnarray}\label{c7}
\begin{array}{c}

M_2(+,-,+,-,+,-)=\,\left\{-4{(2e)}^4\,D(k_3^2)\,N_1(p_1,p_2)N_2(p_1,p_2)\Bigl[(p_1p_2)\sqrt{(k_1p_4)(k_1p_1)}\times\right.
\\
\sqrt{(p_1p_3)}e^{i(\Delta\varphi_{9})}-(p_1p_2)(p_1p_3)\sqrt{(p_3p_4)}-\sqrt{(p_1p_2)(k_1p_2)(k_1p_3)(k_1p_1)(k_1p_4)}e^{i(\Delta\varphi_{10})}+
\\
\left.\sqrt{(p_1p_3)(p_1p_2)(k_1p_2)(k_1p_3)(p_3p_4)}e^{i(\Delta\varphi_{1})}\Bigr]\sqrt{(p_1p_2)}e^{i(\Delta\varphi_{11})}\right\}-\Bigl\{{(p_1\leftrightarrow
p_3)}\Bigr\} e^{i(\Delta\varphi_{12})}-
\\
 \Bigl\{{(p_2\leftrightarrow p_4)}\Bigr\}
e^{i(\Delta\varphi_{13})}+ \Bigl\{{(p_1\leftrightarrow
p_3,p_2\leftrightarrow p_4)}\Bigr\} e^{i(\Delta\varphi_{2+-})},
\end{array}
\end{eqnarray}

 \vspace{0.5cm}

\begin{eqnarray}\label{c8}
\begin{array}{c}

M_2(+,-,-,+,+,-)=\,\left\{\frac{\displaystyle\Bigr[4{(2e)}^4\,D(k_3^2)\,N_1(p_1,p_2)N_2(p_1,p_2)}{\displaystyle(p_1p_4)(p_3p_4)}\left\{\Bigl[(k_2p_1)(k_2p_4)
\times \right. \right.
 \\ \sqrt{(p_3p_4)(p_3p_2)(p_1p_2)}+\Bigl[(p_1p_4)(p_3p_4)-(k_2p_1)(p_3p_4)\Bigr]\sqrt{(p_1p_2)(k_2p_2)(k_2p_4)}e^{i(\Delta\varphi_{14})}+
  \\
  \Bigl[(p_1p_2)(p_1p_4)-(p_1p_2)(k_2p_4)\Bigr]\sqrt{(k_2p_1)(k_2p_3)(p_3p_4)}e^{i(\Delta\varphi_{15})}-
  (p_3p_4)(p_1p_4)(p_1p_2)
  \sqrt{(p_1p_4)}\times \\ \left. e^{i(\Delta\varphi_{16})}\right\}\sqrt{(p_1p_2)(p_1p_4)(p_3p_4)}e^{i(\Delta\varphi_{17})}\Biggr\}+ \Bigl\{{(p_1\leftrightarrow
p_3,p_2\leftrightarrow p_4),k_1\leftrightarrow k_2)}\Bigr\}^{*}
e^{i(\Delta\varphi_{2+-})},
\end{array}
\end{eqnarray}

$$M_2(+,+,+,-,+,-)=0,$$ $$M_2(+,+,-,+,+,-)=0.$$

 \vspace{0.2cm} \large Matrix elements (\ref{c9}) -
 (\ref{c12}) correspond to the third diagram.
Basic polarization states of interacting particles are considered:
\large

\begin{eqnarray}\label{c9}
\begin{array}{c}

M_3(+,-,+,-,+,-)=\,\left\{-4{(2e)}^4\left[\,\frac{\displaystyle
D(k_3^2)\,N_1(p_1,p_2)N_2(p_1,p_2)}{\displaystyle
\Bigl((p_1p_3)+(p_1p_4)+(p_3p_4)\Bigr)}\Bigl[\sqrt{(k_1p_1)(p_1p_4)}\right.\right.+
 \\
 \sqrt{(k_1p_3)(p_3p_4)}e^{i(\Delta\varphi_{18})}\Bigr](p_1p_2)\sqrt{(p_1p_3)(k_2p_1)(k_2p_2)(k_1p_1)}-
 \,\frac{\displaystyle
D(k_3^2)\,N_1(p_1,p_2)N_2(p_1,p_2)}{\displaystyle
\Bigl((p_2p_4)+(p_3p_2)+(p_3p_4)\Bigr)}\times\\
\Bigl[\sqrt{(k_2p_2)(p_3p_2)}+
 \sqrt{(k_2p_4)(p_3p_4)}e^{i(\Delta\varphi_{19})}\Bigr](p_1p_2)
 \sqrt{(p_2p_4)(k_2p_2)(k_1p_2)(k_1p_1)}e^{i(\Delta\varphi_{20})}\Bigr]\times \\  e^{i(\Delta\varphi_{21})}\Biggr\}-
\Bigl\{{(p_1\leftrightarrow p_3)}\Bigr\}e^{i(\Delta\varphi_{12})}
- \Bigl\{{(p_2\leftrightarrow
p_4)}\Bigr\}e^{i(\Delta\varphi_{13})}+ \Bigl\{{(p_1\leftrightarrow
p_3,p_2\leftrightarrow p_4)}\Bigr\}e^{i(\Delta\varphi_{2+-})},
\end{array}
\end{eqnarray}

\vspace{0.5cm}

\begin{eqnarray}\label{c10}
\begin{array}{c}

M_3(+,-,-,+,+,-)=\,\left\{4{(2e)}^4\left[\,\frac{\displaystyle
D(k_3^2)\,N_1(p_1,p_2)N_2(p_1,p_2)}{\displaystyle
\Bigl((p_1p_3)+(p_1p_4)+(p_3p_4)\Bigr)}\Bigl[(k_2p_1)\sqrt{(p_1p_3)}\right.\right.+
 \\
 \sqrt{(k_2p_1)(k_2p_4)(p_3p_4)}e^{i(\Delta\varphi_{22})}\Bigr](p_1p_2)\sqrt{(k_1p_1)(k_1p_2)(p_1p_4)}+
 \,\frac{\displaystyle
D(k_3^2)\,N_1(p_1,p_2)N_2(p_1,p_2)}{\displaystyle
\Bigl((p_2p_4)+(p_3p_2)+(p_3p_4)\Bigr)}\times\\
\Bigl[(k_1p_2)\sqrt{(p_2p_4)}+
 \sqrt{(k_1p_2)(k_1p_3)(p_3p_4)}e^{i(\Delta\varphi_{23})}\Bigr](p_1p_2)\sqrt{(k_2p_1)(k_2p_2)(p_3p_2)} e^{i(\Delta\varphi_{24})}\Bigr]  \times
 \\e^{i(\Delta\varphi_{25})}\Biggr\}+
\Bigl\{{(p_1\leftrightarrow p_3,p_2\leftrightarrow
p_4),k_1\leftrightarrow k_2)}\Bigr\}^{*}
e^{i(\Delta\varphi_{26})},
\end{array}
\end{eqnarray}

\vspace{0.5cm} \npage \large
\begin{eqnarray}\label{c11}
\begin{array}{c}
M_3(+,+,+,-,+,-)=\,\left\{-4{(2e)}^4\,\frac{\displaystyle
D(k_3^2)\,N_1(p_1,p_2)N_2(p_1,p_2)}{\displaystyle
\Bigl((p_2p_4)+(p_3p_2)+(p_3p_4)\Bigr)}\Bigl[\sqrt{(k_2p_1)(p_1p_2)}\right.-
 \\
\sqrt{(k_1k_2)(k_1p_2)}e^{i(\Delta\varphi_{27})}\Bigr](p_1p_2)(p_2p_4)\sqrt{(p_3p_4)(k_2p_1)}e^{i(\Delta\varphi_{28})}\Biggr\}-
\Bigl\{{(p_1\leftrightarrow p_3)}\Bigr\}e^{i(\Delta\varphi_{29})}
-
\\ \Bigl\{{(p_2\leftrightarrow
p_4)}\Bigr\}e^{i(\Delta\varphi_{30})}+ \Bigl\{{(p_1\leftrightarrow
p_3,p_2\leftrightarrow p_4)}\Bigr\}e^{i(\Delta\varphi_{2++})}- \\
\Bigl\{{(k_1\leftrightarrow k_2,p_1\leftrightarrow
p_3)}\Bigr\}e^{i(\Delta\varphi_{29})}e^{i(\Delta\varphi_{2++})}-
\Bigl\{{(k_1\leftrightarrow k_2,p_2\leftrightarrow
p_4)}\Bigr\}e^{i(\Delta\varphi_{30})}e^{i(\Delta\varphi_{2++})}+\\
\Bigl\{{(k_1\leftrightarrow k_2,p_1\leftrightarrow
p_3,p_2\leftrightarrow
p_4)}\Bigr\}e^{i(\Delta\varphi_{2++})}+\Bigl\{{(k_1\leftrightarrow
k_2)}\Bigr\}e^{i(\Delta\varphi_{2++})},
\end{array}
\end{eqnarray}
\vspace{0.5cm}
\begin{eqnarray}\label{c12}
\begin{array}{c}

M_3(+,+,-,+,+,-)=\,\left\{4{(2e)}^4\left[\,\frac{\displaystyle
D(k_3^2)\,N_1(p_1,p_2)N_2(p_1,p_2)}{\displaystyle
\Bigl((p_1p_3)+(p_1p_4)+(p_3p_4)\Bigr)}\Bigl[\sqrt{(k_2p_1)(k_1k_2)}\right.\right.\times
 \\
 \sqrt{(k_1p_2)(p_1p_2)}e^{i(\Delta\varphi_{31})}-(k_1p_2)(p_1p_2)\Bigr](p_1p_4)\sqrt{(p_1p_2)(p_3p_4)}+
 \,\frac{\displaystyle
D(k_3^2)\,N_1(p_3,p_4)N_2(p_3,p_4)}{\displaystyle
\Bigl((p_2p_4)+(p_3p_2)+(p_3p_4)\Bigr)}\\
\Bigl[\sqrt{(k_2p_3)(k_1k_2)(k_1p_4)(p_3p_4)}e^{i(\Delta\varphi_{32})}-(k_2p_3)(p_3p_4)\Bigr](p_1p_4)\sqrt{(p_1p_2)(p_3p_4)}
e^{i(\Delta\varphi_{2++})}\Bigr] e^{i(\Delta\varphi_{28})}\Biggr\}
+ \\
 \Bigl\{{(k_1\leftrightarrow
k_2)}\Bigr\}e^{i(\Delta\varphi_{2++})}.

\end{array}
\end{eqnarray}

\vspace{1cm}
 \large
\noindent The following notations were used in equations
(\ref{c3}) - (\ref{c12}): $(p\;k) = p_0k_0 - \vec{\mathstrut
p}\vec{\mathstrut k}$ -- scalar production of four momentum $p$
and $k$,  $ D(k_3^2)$ -- boson propagator,  where $k_3$ is four
momentum of virtual particle  ($\gamma$ or $Z$),

\vspace{-0.3cm}
 \large
\begin{eqnarray}\label{c13}
\begin{array}{c}

 N_1(p_i,p_k) =
[16(p_ip_k)(p_ik_1)(p_kk_1)]^{-1/2}, \\ \\ N_2(p_i,p_k) =
[16(p_ip_k)(p_ik_2)(p_kk_2)]^{-1/2}.
\end{array}
\end{eqnarray}
\vspace{0.1cm}

\noindent
$M_(\lambda_1,\lambda_2,\lambda_3,\lambda_4,\lambda_5,\lambda_6)$
denotes the matrix element,  where  $\lambda_{1(2)}\;$ corresponds
to polarization of photon with four momentum $k_{1(2)}$,
$\lambda_{3,4,5,6}\;-$ polarization of lepton with four momentum
$p_{1,2,3,4}$\;.

The list of all used phase factors is presented in appendix.

 \large Squared matrix element  for each certain
polarization state of interacting particles has the form: \large
$$|M(+,+,+,-,+,-)|^2 = |M_1(+,+,+,-,+,-) + M_2(+,+,+,-,+,-) +
M_3(+,+,+,-,+,-)|^2.$$

\large
 Using symmetry  one can obtain the following relations:

 \vspace{-0.1cm}

\large
\begin{eqnarray}\label{c14}
\begin{array}{c}

\vspace{0.2cm} \,
|M(+,+,-,+,-,+)|^2,=|M(+,+,+,-,+,-)|^2_{p_1\leftrightarrow
p_2,p_3\leftrightarrow p_4},
\\ \vspace{0.2cm}
 |M(+,-,+,-,-,+)|^2 =
|M(+,-,-,+,+,-)|^2_{p_1\leftrightarrow p_3,p_2\leftrightarrow
p_4}, \\ \vspace{0.2cm}   |M(+,+,+,-,-,+)|^2 =
|M(+,+,-,+,+,-)|^2_{p_1\leftrightarrow p_3,p_2\leftrightarrow
p_4},
\\ \vspace{0.2cm} \!\!\!\!\!\!\!\!\!\!\!\!\!\!|M(+,-,-,+,-,+)|^2 =
|M(+,-,+,-,+,-)|^2_{k_1\leftrightarrow k_2}, \\ \vspace{0.2cm}
\!\!\!\!\!\!\!\!\!\!\!\!\!\!|M(+,-,+,+,-,-)|^2
=|M(+,-,-,+,+,-)|^2_{p_1\leftrightarrow p_3},\\ \vspace{0.2cm}
\!\!\!\!\!\!\!\!\!\!\!\!\!\!|M(+,-,-,-,+,+)|^2
=|M(+,-,-,+,+,-)|^2_{p_2\leftrightarrow p_4}, \\ \vspace{0.2cm}
 \!\!\!\!\!\!\!\!\!\!\!\!\!\!|M(+,+,+,+,-,-)|^2
=|M(+,+,-,+,+,-)|^2_{p_1\leftrightarrow p_3},  \\ \vspace{0.2cm}
\!\!\!\!\!\!\!\!\!\!\!\!\!\!|M(+,+,-,-,+,+)|^2
=|M(+,+,-,+,+,-)|^2_{p_2\leftrightarrow p_4},
\end{array}
\end{eqnarray}
\npage \large
\begin{eqnarray}\label{c15}
\begin{array}{c}
\vspace{0.1cm}
  |M(-,+,-,+,-,+)|^2=|M(+,-,+,-,+,-)|^2, \vspace{0.1cm} \\ \vspace{0.1cm}
|M(-,-,-,+,+,-)|^2=|M(+,+,+,-,-,+)|^2,
 \vspace{0.1cm} \\ \vspace{0.1cm}
|M(-,-,+,-,+,-)|^2=|M(+,+,-,+,-,+)|^2, \vspace{0.1cm} \\
\vspace{0.1cm} |M(-,+,+,-,-,+)|^2=|M(+,-,-,+,+,-)|^2,
\vspace{0.1cm} \\ \vspace{0.1cm}
|M(-,-,-,+,-,+)|^2=|M(+,+,+,-,+,-)|^2, \vspace{0.1cm}
\\\vspace{0.1cm} |M(-,+,-,+,+,-)|^2=|M(+,-,+,-,-,+)|^2,
\vspace{0.1cm} \\\vspace{0.1cm}
|M(-,-,+,-,-,+)|^2=|M(+,+,-,+,+,-)|^2, \vspace{0.1cm}
\\\vspace{0.1cm} |M(-,+,+,-,+,-)|^2=|M(+,-,-,+,-,+)|^2,
\vspace{0.1cm} \\ \vspace{0.1cm}
|M(-,+,-,-,+,+)|^2=|M(+,-,+,+,-,-)|^2, \vspace{0.1cm}
\\\vspace{0.1cm} |M(-,+,+,+,-,-)|^2=|M(+,-,-,-,+,+)|^2,
\vspace{0.1cm} \\\vspace{0.1cm}
|M(-,-,-,+,+,-)|^2=|M(+,+,+,-,-,+)|^2, \vspace{0.1cm}
\\\vspace{0.1cm} |M(-,-,+,+,-,-)|^2=|M(+,+,-,-,+,+)|^2.
\end{array}
\end{eqnarray}
to simplify real calculations.
 \section { Conclusion} $$ $$

{

Matrix element and differential cross section  of process
$\gamma\gamma\rightarrow\;l^+l^-l^+l^-$   are calculated using
helicity amplitude method at every possible polarization state.
Obtained formulas don't contain any bispinor. It allows
 to square and   integrate these amplitudes effectively.
 Numerical calculation of differential and total cross sections under
different kinematics cuts as well as the main problems of
integration will be discussed in the next paper.

\npage

$$ $$\section{ Appendix} $$ $$ \large \vspace{-0.5cm} { \bf The
whole list of used phase factors:} \vspace{0.2cm}
$$e^{i\Delta\varphi_1}=\frac{Tr((\hat{k}_1\hat{p}_2\hat{p}_1\hat{p}_3)(1-\gamma_5))}{8\sqrt{(p_1p_3)(p_1p_2)(k_1p_2)(k_1p_3)}};$$

$$e^{i\Delta\varphi_2}=\frac{((p_1p_3)(k_1p_2)-(p_2p_3)(k_1p_1))Tr((\hat{k}_1\hat{p}_2\hat{p}_1\hat{p}_3)(1-\gamma_5))-((p_1p_2)(k_1p_3))Tr((\hat{k}_1\hat{p}_2\hat{p}_3\hat{p}_1)(1-\gamma_5))}{8(p_1p_3)(p_1p_2)(k_1p_2)(k_1p_3)};$$

$$e^{i\Delta\varphi_3}=\frac{Tr((\hat{p}_3\hat{p}_2\hat{p}_1\hat{p}_4)(1-\gamma_5))}{8\sqrt{(p_1p_4)(p_3p_2)(p_1p_2)(p_3p_4)}}\times(\varepsilon_1^{-}(p_3,p_2)\varepsilon_1^{+}(p_1,p_2))\times(\varepsilon_2^{+}(p_1,p_4)\varepsilon_2^{-}(p_3,p_4));$$

$$e^{i\Delta\varphi_4}=\frac{Tr((\hat{p}_3\hat{p}_2\hat{p}_1\hat{p}_4)(1-\gamma_5))}{8\sqrt{(p_1p_4)(p_3p_2)(p_1p_2)(p_3p_4)}}\times(\varepsilon_1^{-}(p_1,p_4)\varepsilon_1^{+}(p_1,p_2))\times(\varepsilon_2^{+}(p_3,p_2)\varepsilon_2^{-}(p_3,p_4));$$

$$e^{i\Delta\varphi_{1+,-}}=(\varepsilon_1^{-}(p_3,p_4)\varepsilon_1^{+}(p_1,p_2))\times(\varepsilon_2^{+}(p_1,p_2)\varepsilon_2^{-}(p_3,p_4));$$
$$e^{i\Delta\varphi_{2+,+}}=(\varepsilon_1^{-}(p_3,p_4)\varepsilon_1^{+}(p_1,p_2))\times(\varepsilon_2^{-}(p_1,p_2)\varepsilon_2^{+}(p_3,p_4));$$

$$e^{i\Delta\varphi_5}=\frac{Tr((\hat{k}_1\hat{p}_1\hat{p}_2\hat{p}_3)(1-\gamma_5))}{8\sqrt{(p_1p_2)(p_3p_2)(k_1p_3)(k_1p_1)}};$$

$$e^{i\Delta\varphi_6}=\frac{((p_1p_2)(k_1p_3)-(p_1p_3)(k_1p_2))Tr((\hat{k}_1\hat{p}_1\hat{p}_2\hat{p}_3)(1-\gamma_5))-((p_2p_3)(k_1p_1))Tr((\hat{k}_1\hat{p}_2\hat{p}_1\hat{p}_3)(1-\gamma_5))}{8(p_1p_2)(p_2p_3)(k_1p_1)(k_1p_3)};$$

$$e^{i\Delta\varphi_7}=\frac{Tr((\hat{p}_3\hat{p}_2\hat{p}_1\hat{p}_4)(1-\gamma_5))}{8\sqrt{(p_1p_4)(p_3p_2)(p_1p_2)(p_3p_4)}}\times\frac{Tr((\hat{p}_3\hat{p}_2\hat{p}_1\hat{p}_4)(1-\gamma_5))}{8\sqrt{(p_1p_4)(p_3p_2)(p_1p_2)(p_3p_4)}}\times
e^{i\Delta\varphi_{1+-}};$$

$$e^{i\Delta\varphi_8}=\frac{Tr((\hat{p}_4\hat{p}_3\hat{p}_2\hat{p}_1)(1-\gamma_5))}{8\sqrt{(p_3p_4)(p_3p_2)(p_1p_4)(p_1p_2)}}\times(\varepsilon_1^{-}(p_3,p_2)\varepsilon_1^{+}(p_1,p_2))\times(\varepsilon_2^{-}(p_1,p_4)\varepsilon_2^{+}(p_3,p_4));$$

$$e^{i\Delta\varphi_9}=\frac{Tr((\hat{k}_1\hat{p}_4\hat{p}_3\hat{p}_1)(1-\gamma_5))}{8\sqrt{(p_1p_3)(p_3p_4)(k_1p_4)(k_1p_1)}};$$

$$e^{i\Delta\varphi_{10}}=\frac{((p_3p_4)(k_1p_1)-(p_1p_4)(k_1p_3))Tr((\hat{p}_2\hat{k}_1\hat{p}_3\hat{p}_1)(1-\gamma_5))-((p_1p_3)(k_1p_3))Tr((\hat{p}_2\hat{k}_1\hat{p}_4\hat{p}_1)(1-\gamma_5))}{8(p_1p_3)\sqrt{(p_1p_2)(k_1p_2)(k_1p_3)(k_1p_1)(k_1p_4)(p_3p_4)}};$$

$$e^{i\Delta\varphi_{11}}=(\varepsilon_2^{+}(p_3,p_4)\varepsilon_2^{-}(p_1,p_2));$$

$$e^{i\Delta\varphi_{12}}=\frac{Tr((\hat{p}_3\hat{p}_2\hat{p}_1\hat{p}_4)(1-\gamma_5))}{8\sqrt{(p_1p_4)(p_3p_2)(p_1p_2)(p_3p_4)}}\times(\varepsilon_1^{-}(p_3,p_2)\varepsilon_1^{+}(p_1,p_2))\times(\varepsilon_2^{+}(p_3,p_2)\varepsilon_2^{-}(p_1,p_2));$$

$$e^{i\Delta\varphi_{13}}=\frac{Tr((\hat{p}_3\hat{p}_2\hat{p}_1\hat{p}_4)(1-\gamma_5))}{8\sqrt{(p_1p_4)(p_3p_2)(p_1p_2)(p_3p_4)}}\times(\varepsilon_1^{-}(p_1,p_4)\varepsilon_1^{+}(p_1,p_2))\times(\varepsilon_2^{+}(p_1,p_4)\varepsilon_2^{-}(p_1,p_2));$$

$$e^{i\Delta\varphi_{2+,-}}=(\varepsilon_1^{-}(p_3,p_4)\varepsilon_1^{+}(p_1,p_2))\times(\varepsilon_2^{+}(p_3,p_4)\varepsilon_2^{-}(p_1,p_2));$$
$$e^{i\Delta\varphi_{2+,+}}=(\varepsilon_1^{-}(p_3,p_4)\varepsilon_1^{+}(p_1,p_2))\times(\varepsilon_2^{-}(p_3,p_4)\varepsilon_2^{+}(p_1,p_2));$$
\npage\large
$$e^{i\Delta\varphi_{14}}=\frac{Tr((\hat{p}_2\hat{k}_2\hat{p}_4\hat{p}_3)(1+\gamma_5))}{8\sqrt{(p_3p_4)(p_3p_2)(k_2p_2)(k_2p_4)}};\;\;
e^{i\Delta\varphi_{15}}=\frac{Tr((\hat{k}_2\hat{p}_3\hat{p}_2\hat{p}_1)(1+\gamma_5))}{8\sqrt{(p_3p_2)(p_1p_2)(k_2p_3)(k_2p_1)}};$$

$$e^{i\Delta\varphi_{16}}=\frac{Tr((\hat{p}_4\hat{p}_3\hat{p}_2\hat{p}_1)(1+\gamma_5))}{8\sqrt{(p_1p_4)(p_3p_4)(p_1p_2)(p_3p_2)}};\;\;e^{i\Delta\varphi_{17}}=\frac{Tr((\hat{p}_4\hat{p}_1\hat{p}_2\hat{p}_3)(1-\gamma_5))}{8\sqrt{(p_3p_2)(p_1p_2)(p_3p_4)(p_1p_4)}}\times(\varepsilon_2^{+}(p_3,p_4)\varepsilon_2^{-}(p_1,p_2));$$

$$e^{i\Delta\varphi_{18}}=\frac{Tr((\hat{p}_1\hat{k}_1\hat{p}_3\hat{p}_4)(1-\gamma_5))}{8\sqrt{(k_1p_1)(k_1p_3)(p_3p_4)(p_1p_4)}};\;\;e^{i\Delta\varphi_{19}}=\frac{Tr((\hat{p}_2\hat{k}_2\hat{p}_4\hat{p}_3)(1-\gamma_5))}{8\sqrt{(k_2p_2)(k_2p_4)(p_3p_4)(p_3p_2)}};$$

$$e^{i\Delta\varphi_{20}}=\frac{Tr((\hat{p}_2\hat{k}_1\hat{p}_1\hat{p}_3)(1-\gamma_5))}{8\sqrt{(k_1p_2)(k_1p_1)(p_1p_3)(p_3p_2)}}\times\frac{Tr((\hat{p}_4\hat{p}_2\hat{k}_2\hat{p}_1)(1-\gamma_5))}{8\sqrt{(p_2p_4)(k_2p_2)(k_2p_1)(p_1p_4)}};$$

$$e^{i\Delta\varphi_{21}}=\frac{((p_1p_4)Tr((\hat{k}_2\hat{p}_2\hat{p}_3\hat{p}_1)(1-\gamma_5))-(p_1p_3)Tr((\hat{k}_2\hat{p}_2\hat{p}_4\hat{p}_1)(1-\gamma_5)))}{8\sqrt{(k_2p_2)(p_1p_2)(p_1p_4)(p_3p_4)(p_1p_3)(k_2p_1)}}\times^{i\Delta\varphi_{11}};$$

$$e^{i\Delta\varphi_{22}}=\frac{Tr((\hat{p}_1\hat{p}_3\hat{p}_4\hat{k}_2)(1-\gamma_5))}{8\sqrt{(k_2p_1)(k_2p_4)(p_3p_4)(p_1p_3)}};\;\;e^{i\Delta\varphi_{23}}=\frac{Tr((\hat{p}_4\hat{p}_2\hat{k}_1\hat{p}_3)(1+\gamma_5))}{8\sqrt{(k_1p_2)(k_1p_3)(p_2p_4)(p_3p_4)}};$$

$$e^{i\Delta\varphi_{24}}=\frac{Tr((\hat{p}_4\hat{p}_2\hat{k}_2\hat{p}_1)(1-\gamma_5))}{8\sqrt{(p_2p_4)(k_2p_2)(k_2p_1)(p_1p_4)}}\times\frac{Tr((\hat{p}_2\hat{p}_3\hat{p}_1\hat{k}_1)(1+\gamma_5))}{8\sqrt{(p_3p_2)(p_1p_3)(k_1p_2)(k_1p_1)}};$$

$$e^{i\Delta\varphi_{25}}=-\frac{Tr((\hat{p}_4\hat{p}_1\hat{p}_2\hat{p}_3)(1-\gamma_5))}{8\sqrt{(p_1p_4)(p_1p_2)(p_3p_2)(p_3p_4)}}\times\frac{Tr((\hat{k}_1\hat{p}_2\hat{p}_3\hat{p}_1)(1+\gamma_5))}{8\sqrt{(k_1p_2)(p_3p_2)(p_1p_3)(k_1p_1)}}\times
e^{i\Delta\varphi_{11}};$$

$$e^{i\Delta\varphi_{26}}=-\frac{Tr((\hat{p}_4\hat{p}_1\hat{p}_3\hat{k}_2)(1+\gamma_5))}{8\sqrt{(p_1p_4)(p_1p_3)(k_2p_4)(k_2p_3)}}\times\frac{Tr((\hat{k}_1\hat{p}_2\hat{p}_3\hat{p}_1)(1-\gamma_5))}{8\sqrt{(p_3p_2)(p_1p_3)(k_1p_2)(k_1p_1)}}\times
e^{i\Delta\varphi_{2+,-}};$$

$$e^{i\Delta\varphi_{27}}=\frac{Tr((\hat{p}_1\hat{k}_2\hat{k}_1\hat{p}_2)(1-\gamma_5))}{8\sqrt{(k_2p_1)(k_1k_2)(k_1p_2)(p_1p_2)}};\;\;e^{i\Delta\varphi_{28}}=(\varepsilon_2^{-}(p_3,p_4)\varepsilon_2^{+}(p_1,p_2));$$
$$e^{i\Delta\varphi_{29}}=\frac{Tr((\hat{p}_4\hat{p}_3\hat{p}_2\hat{p}_1)(1-\gamma_5))}{8\sqrt{(p_3p_4)(p_3p_2)(p_1p_4)(p_1p_2)}}\times(\varepsilon_1^{-}(p_3,p_2)\varepsilon_1^{+}(p_1,p_2))\times(\varepsilon_2^{-}(p_3,p_2)\varepsilon_2^{+}(p_1,p_2));$$

$$e^{i\Delta\varphi_{30}}=\frac{Tr((\hat{p}_4\hat{p}_3\hat{p}_2\hat{p}_1)(1-\gamma_5))}{8\sqrt{(p_3p_4)(p_3p_2)(p_1p_4)(p_1p_2)}}\times(\varepsilon_1^{-}(p_1,p_4)\varepsilon_1^{+}(p_1,p_2))\times(\varepsilon_2^{-}(p_1,p_4)\varepsilon_2^{+}(p_1,p_2));$$

$$e^{i\Delta\varphi_{31}}=\frac{Tr((\hat{k}_2\hat{k}_1\hat{p}_2\hat{p}_1)(1+\gamma_5))}{8\sqrt{(k_1k_2)(k_1p_2)(p_1p_2)(k_1p_2)}};\;\;e^{i\Delta\varphi_{32}}=\frac{Tr((\hat{p}_4\hat{p}_3\hat{k}_2\hat{k}_1)(1+\gamma_5))}{8\sqrt{(k_1k_2)(p_3p_4)(k_2p_3)(k_1p_4)}}.$$

Here the following notations are used:

$$\varepsilon_1^\mp(p_i,p_k)\varepsilon_1^\pm(p_1,p_2)=-N_1(p_i,p_k)N_1(p_1,p_2)Tr(p_ip_kk_1p_1p_2k_1(1\mp\gamma_5)),$$
$$\varepsilon_2^\pm(p_i,p_k)\varepsilon_2^\mp(p_3,p_4)=-N_2(p_i,p_k)N_2(p_3,p_4)Tr(p_ip_kk_2p_3p_4k_2(1\pm\gamma_5)),$$
$N_1(p_i,p_k)$ and $N_2(p_i,p_k)$ have been defined by equation
(\ref{c13}).

\npage

\end{document}